\documentclass[12pt,preprint]{aastex}

\usepackage{ulem}

\newcommand{\mearth}{\rm M_\oplus}

\usepackage{xcolor}
\slugcomment{to appear in ApJL}

\shorttitle{Blocking the inward migration of Super-Earths}
\shortauthors{Izidoro et al.}

\begin{document}


\title{Gas giant planets as dynamical barriers to inward-migrating super-Earths}


\author{Andr{\'e} Izidoro\altaffilmark{1,2,3}, Sean N. Raymond\altaffilmark{3,4}, Alessandro Morbidelli\altaffilmark{1},}

\author{Franck Hersant\altaffilmark{3,4} and Arnaud Pierens\altaffilmark{3,4}}



\altaffiltext{1}{University of Nice-Sophia Antipolis, CNRS, Observatoire de la C{\^o}te d'Azur, Laboratoire Lagrange, BP4229, F-06304 NICE Cedex 4, France.}
\altaffiltext{2}{Capes Foundation, Ministry of Education of Brazil, Bras{\'i}lia/DF 70040-020, Brazil.}
\altaffiltext{3}{Univ. Bordeaux, Laboratoire d'Astrophysique de Bordeaux, UMR 5804, F-33270 Floirac, France}
\altaffiltext{4}{CNRS, Laboratoire d'Astrophysique de Bordeaux, UMR 5804, F-33270 Floirac, France}

\begin{abstract}
Planets of 1-4 times Earth's size on orbits shorter than 100 days exist around 30-50\% of all Sun-like stars. In fact, the Solar System is particularly outstanding in its lack of ``hot super-Earths'' (or ``mini-Neptunes'').  These planets -- or their building blocks -- may have formed on wider orbits and migrated inward due to interactions with the gaseous protoplanetary disk. Here, we use a suite of dynamical simulations to show that gas giant planets act as barriers to the inward migration of super-Earths initially placed on more distant orbits.  Jupiter's early formation may have prevented Uranus and Neptune (and perhaps Saturn's core) from becoming hot super-Earths. Our model predicts that the populations of hot super-Earth systems and Jupiter-like planets should be anti-correlated: gas giants (especially if they form early) should be rare in systems with many hot super-Earths. Testing this prediction will constitute a crucial assessment of the validity of the migration hypothesis for the origin of close-in super-Earths.
\end{abstract}


\section{Introduction}

According to the standard core accretion model (Pollack et al., 1996), gas giant planets form in a few million years in gas-dominated disks.  A solid core of a few to dozens of Earth masses grows then accretes a thick gaseous envelope. It is believed that giant planet cores form preferentially in a region of the protoplanetary disk where the radial drift speed of dust, pebbles and small planetesimals is slowed, creating a localized enhancement in the density of solid material (Johansen et al., 2009). One favorable location for this to happen is at the snowline (Kokubo \& Ida, 1998; Bitsch et al., 2014a,b), where water condenses as ice (Lecar et al., 2006). Planet formation models tend to produce not one but multiple planet cores beyond the snowline (Kokubo \& Ida, 1998; Lambrechts et al., 2014). However, observational evidence suggests that not all emerging planetary cores are able to accrete enough gas to become gas giant planets. Indeed, Uranus and Neptune may be thought of as ``failed'' giant planet cores, and Neptune-sized exoplanets have been found to be far more abundant than Jupiter-sized ones (Mayor et al., 2011; Howard et al., 2010).  

Super-Earths are extremely abundant (Mayor et al., 2011; Howard et al., 2012; here ``super-Earths'' are defined as all planets between 1 and 20 Earth masses or 1 and 4 Earth radii).  An extraordinary number of planetary systems with multiple super-Earths have been found (Mayor et al., 2011; Howard et al., 2010; 2012;  Fressin et al., 2013;  Marcy et al., 2014).  These planets' orbits are typically found in tightly-packed configurations and located much closer to their stars than expected, well interior to the primordial snowline (Lissauer et al., 2011a).  The origin of these systems is an open debate (Raymond et al., 2008; 2014).  There are two competing models: super-Earths either formed in situ (Raymond et al., 2008; Hansen \& Murray, 2012; 2013; Chiang \& Laughlin, 2013) or migrated towards the central star from the outer regions of their proto-planetary disks (Terquem \& Papaloizou, 2007; McNeil \& Nelson, 2010; Ida \& Lin, 2010;  Raymond \& Cossou, 2014; Cossou et al., 2014). Unlike the inward migration model, the in-situ formation scenario requires either congenital high disk mass or effective drift of solid precursors ($\sim$cm-to-m size objects) to the inner regions of the disk creating enhanced surface densities of solids (Hansen, 2014; Chatterjee \& Tan, 2014; Boley \& Ford, 2013).

Super-Earths are so abundant that any model for their origin must be very efficient.  The key question then becomes: {\it why are there no hot super-Earths in our Solar System?}

Here we propose that an interplay between growing gas giants and migrating super-Earths can solve this mystery.  If super-Earths (or their constituent embryos) form in the outer disk and migrate towards the central star some of these objects must on occasion become giant planetary cores (see Cossou et al 2014).  If the innermost super-Earth in a young planetary system grows into a gas giant then the later dynamical evolution of the system is drastically changed.  The giant planet blocks the super-Earths' inward migration.  In this context, Jupiter may have prevented Uranus and Neptune -- and perhaps Saturn's core -- from migrating inward and becoming super-Earths.  This model -- which we test below with numerical simulations -- predicts an anti-correlation between the populations of giant exoplanets and super-Earths that will be tested in coming datasets.  No such anti-correlation is expected from the in-situ model of super-Earth formation (Schlaufman, 2014).  This therefore represents a testable way to differentiate between models. 

We first test the concept of gas giants as barriers to inward-migrating super-Earths or planetary embryos in section 2.  We then analyze the expected observational consequences and make specific testable predictions in section 3.  We also discuss the uncertainties and how much can be inferred from upcoming observations.  

\section{The model: gas giants as type I migration barriers}

We ran a suite of N-body simulations to test whether giant planets act as dynamical barriers for inward migrating super-Earths. Our simulations represented the middle stages of planet formation in gas-dominated circumstellar disks.  The simulations included a population of super-Earths initially located beyond the orbit of one or two fully-formed gas giants.  Our code incorporated the relevant planet-disk interactions, calibrated to match complex hydrodynamical simulations.

A Jupiter-mass planet opens a gap in the protoplanetary disk and type II migrates inward (Lin \& Papaloizou, 1986).  Type-II migration of gas giants is significantly slower than the type-I migration of super-Earths (for super-Earths of $\sim$3 Earth masses or larger for traditional disk parameters). Of course, if there is enough time a giant planet would migrate close to the star and any super-Earth beyond its orbit as well. However, most of giant planets do not appear to have this fate. The vast majority of giant exoplanets around FGK stars are found at a few AU from their host stars (Udry \& Santos, 2007; Mayor et al 2011). Only about 0.5-1\% of the sun-like stars host hot-Jupiters (Cumming et al.2008; Howard et al. 2010, 2012; Wright et al., 2012). However, radial velocity and microlensing surveys hint that the abundance of long-period gas giants is of least ∼15\% (Mayor et al 2011; Gould et al 2010). This dissemblance may be a consequence of the inside-out photo-evaporation of the disk, which creates an inner very low-density cavity in the protoplanetary disk that stops inward migration (Alexander \& Pascucci, 2012). In the case of the solar system, the proximity and mass ratio of Jupiter and Saturn may have prevented the inward migration of the giant planets (Morbidelli \& Crida, 2007) or even forced their outward migration (Walsh et al., 2011). In our simulations we assume for simplicity that the giant planet does not migrate, as a proxy of all cases in which there is a differential migration of super-Earths towards the giant planet(s).

In our simulations the central star was taken to be Sun-like (we note that the relevant dynamics are only very weakly stellar mass-dependent). Because the real structures of protoplanetary disks may be quite diverse, we performed simulations exploring a broad swath of parameter space.  We tested four different disk surface density profiles and a range in the number (1, 2, 3, 5, 10 or 20) of and total mass in super-Earths. In all we considered 48 different simulation set-ups.  For each set-up we ran 100 simulations with slightly different initial orbits for the super-Earths. 

\subsection{Numerical Simulations} 

We performed a total of 4800 simulations using the Symba integrator (Duncan et al., 1998) using a 3-day timestep. We modified the code to include the effects of type-I migration and tidal damping acting on super-Earths. Our simulations start with one or two giant planets orbiting beyond 3.5 AU from the central star and a population of super-Earths distributed beyond the orbit of the giant planet(s).  Super-Earths are separated from each other by 5 to 10 mutual Hill radii (eg. Kokubo \& Ida, 1998). The mutual Hill radius is defined as
\begin{equation}
{\rm R_{h,m}=\left( \frac{M_{SE}}{12M_{\odot}}\right)^{\frac{1}{3}}(a_1+a_2)} 
\end{equation}
where ${\rm M_{SE}}$ and ${\rm M_{\odot}}$ are the individual mass of the super-Earths and the central star, respectively. ${\rm a_1}$ and ${\rm a_2}$ are the semi-major axes of any two adjacent super-Earths. Initially,
 eccentricities of the super-Earths and giant planets eccentricities and inclinations are set to be zero. Orbital inclinations of the super-Earths were initially chosen randomly from 0.001 to 0.01 degrees. The argument of pericenter and longitutude of ascending node of the super-Earths are randomly selected between 0 and 360 degrees.

Collisions were treated as inelastic mergers that conserved linear momentum.  Objects were removed from the system if they strayed beyond 100 AU from the central star. In simulations with both Jupiter and Saturn, to ensure the dynamical stability between the two giant planets, the eccentricities of the latter planets are artificially damped if they increase to beyond those values observed in hydrodynamical simulations ($\lesssim$ 0.03-0.05).

\subsubsection{Disk of Gas}

Both the speed and direction of type I migration are sensitive to the properties of the gaseous disk.  Numerical studies have shown that in radiative disks planetary embryos can migrate outward (e.g. Paardekooper \& Papaloizou, 2008; Kley \& Crida, 2008; Paardekooper et al., 2010; 2011). However, outward migration is only possible for a limited range of planetary masses (e.g. Bitsch et al., 2014a). Migration is directed inwards in the regions of the disk where viscous heating is a sufficiently weak heat source.  Because the surface density of a disk decays with time, the region of outward migration shifts inward and eventually disappears. Thus, all low-mass planets eventually migrate inwards (Lyra et al., 2010; Bitsch et al., 2014a; Cossou et al 2014). We focus on this later stage and assume that super-Earths migrate inward at the isothermal Type-I rate (e.g., Tanaka et al 2002).

We used hydrodynamical simulations to obtain surface density profiles of disks with embedded planets.   Figure 1 shows the gas profiles used in our simulations. Disk A contains fully-formed Jupiter and Saturn at 3.5 and 5.0 AU, respectively. Away from the gap its surface density  is within a factor of 2 (larger) than a minimum-mass disk  Disk B initially is identical to disk A but only includes Jupiter at 3.5 AU. The gas density in disks A/B is consistent with inward migration of super-Earths if the viscosity is sufficiently low, giving a stellar accretion rate smaller than {\rm $10^{-8}$} solar masses per year (Bitsch et al., 2014a). The outer part of disk C (beyond the orbit of the giant planet) is identical to that of disk B, but the inner region is depleted to artificially mimic the opening of an inner cavity in the disk due to photo-evaporation of gas.  In practice, we calculated the inner part of disk C from disk B by applying the following rescale function: 
\begin{equation}
{\rm S(r) = \left[ exp\left(2 -\frac{1}{r}\right)\right]  ^3}
\end{equation}
where ${\rm r}$ is the heliocentric distance. Disk D is a global low surface density disk representing a starving disk approaching its end of life. We generated disk D from disk B by simply dividing the surface density by a factor of $\sim20$.  

We read the different gas disk profiles into our N-body code to calculate synthetic forces appropriate for planet-disk interactions. We applied the formulae from Cresswell \& Nelson (2006, 2008) to calculate accelerations to simulate both type I migration and damping of orbital eccentricity and inclination (see Eqs. 10-16 in Izidoro et al. 2014 for more details). These forces were applied to the super-Earths. For simplicity, we assume a  aspect ratio for the disk given by $\simeq 0.045r^{0.25}$, where r is the heliocentric distance.

We mimicked  the gas disk's dissipation as a global exponential decay for the gas surface density, given by ${\rm exp(-t/\tau_{gas})}$, where ${\rm t}$ is the time and ${\rm \tau_{gas}}$ is the gas dissipation timescale.  In our simulations ${\rm \tau_{gas}}$ was set to 1 Myr and the remaining gas was assumed to be instantly dissipated at $t=3$~Myr.

\begin{figure}[h]
\centering
\includegraphics[scale=1.5]{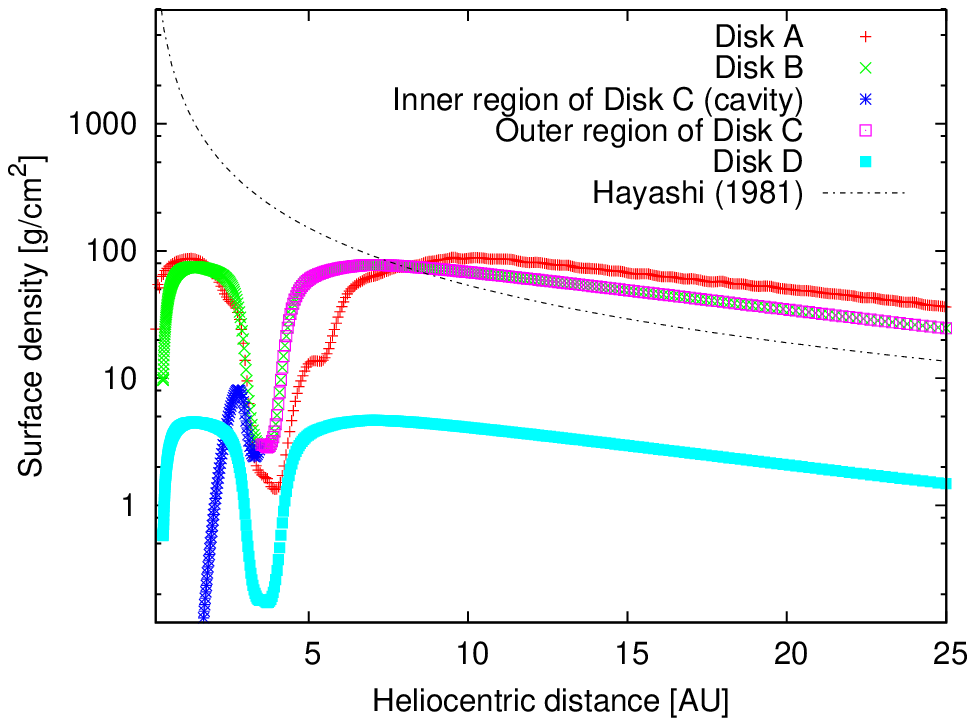}
\vspace{1cm}
\caption{Gap disk profiles from hydrodynamical simulations used to represent the protoplanetary disk in our N-body simulations. The red points (disk A) represent a disk used in simulations modeling the formation of our solar system (Masset \& Snellgrove, 2001; Morbidelli \& Crida, 2007; Walsh et al., 2011). The green points represent an analogous disk that only includes a Jupiter-mass planet at 3.5 AU.  Blue and magenta points represent a disk with an inner cavity opened due to photoevaporation (disk C). The cyan points represent a starving disk which has a very low surface density (disk D). Both disks C and D have a Jupiter-mass planet at 3.5 AU from the central star. For comparison purposes we also show the disk surface density in the traditional minimum mass solar nebula (Hayashi, 1981).}
\end{figure}

\subsection{Simulation Outcomes}

Figure 2 shows the evolution of two characteristic simulations containing a Jupiter-mass planet located at $\sim$3.5 AU.  In both simulations super-Earths migrate inward and pile up just exterior to the gas giant.  Continued migration brings the super-Earths into a more and more compact dynamical configuration (Thommes, 2005). Super-Earths are frequently captured in mean motion resonance with the giant planet and with each other, but eventually come so close to each other as to become unstable.  This leads to a chaotic phase of close encounters and gravitational scattering among super-Earths.  There are several possible outcomes of each instance of this phase.  The super-Earths may collide and merge.  One super-Earth may be ejected from the system into interstellar space.  Ejection tends to happen when a super-Earth is scattered onto an orbit that approaches, but does not cross, the orbit of the giant planet.  Occasionally, a close encounter with the giant planet scatters a super-Earth interior to the giant planet's orbit.  When this happens there are two different pathways for its subsequent evolution: 1) the super-Earth may be scattered back outward and (usually) ejected from the system (see the event at 2.2 Myr in Fig. 1, top panel); or 2) the super-Earth may undergo enough damping of its orbital eccentricity from the gas disk to become dynamically decoupled from the giant planet. When this happens, the super-Earth survives inside the orbit of the giant planet.  Eventually the super-Earth type-I migrates interior to 1 AU (e.g. Figure 2, lower panel). We call planets that cross the orbit of the innermost giant planet and survive in the inner disk ``jumpers''. In the meantime, episodes of dynamical instability continue among the super-Earths that remain beyond the giant planet until the system reaches a final stable/resonant dynamical configuration with few super-Earths surviving (some of which grew in mass by mutual collisions).

The occurrence rate of jumpers is a measure of the strength of the giant planet's dynamical barrier to inward migration (see Figure 3).  The ``jump rate'' is a function of both the properties of the disk and of the number and masses of migrating super-Earths.  In the set-up that is typically assumed for our solar system (``Jupiter and Saturn -- Disk A'' from Fig 2) and with 30 Earth masses in super-Earths, jumpers get past the giant planets in less than 20\% of simulations.  The highest jump rate is for systems with 5-10 migrating super-Earths; this peak is the result of a competition between faster migration for more massive super-Earths and the probability of instability, which is higher for systems with a larger number super-Earths.  For the same disk but with $60 \mearth$ in embryos the jump rate is even higher, with up to 40\% of simulations having jumpers.   The same pattern holds for a range of disk structures, with very low jump rates for systems with fewer than 5-10 super-Earths.  There is a higher rate of jumpers for the single giant planet case because a jumper must pass across the orbits of two giant planets instead of just one.  

Simulations in the disk with an inner cavity (disk C) show a smaller fraction of jumpers.  Tidal damping of a scattered planet's eccentricity in the inner disk is inefficient due to the low gas density so it is difficult for a temporary jumper to dynamically decouple from the giant planet. Finally, simulations in the very low density disk (disk D) have an even smaller frequency of jumpers.  In disk D the jump rate shows no dependence on the initial number or total mass of the super-Earths.  This is a direct consequence of the global low density of the disk, in which a body crossing the orbit of a giant planet can not decouple dynamically from the gas giant and is ejected.

Our simulations probably overestimate of the jump rate, for several reasons.  First, our assumption of isothermal type-I migration for super-Earths and no type-II migration for the gas giants overestimates the differential migration between the two populations.  Second, our assumed equal-mass super-Earths produce stronger instabilities than for super-Earths with unequal masses (Raymond et al., 2010). 

\begin{figure}[!h]
\centering
\includegraphics[scale=1]{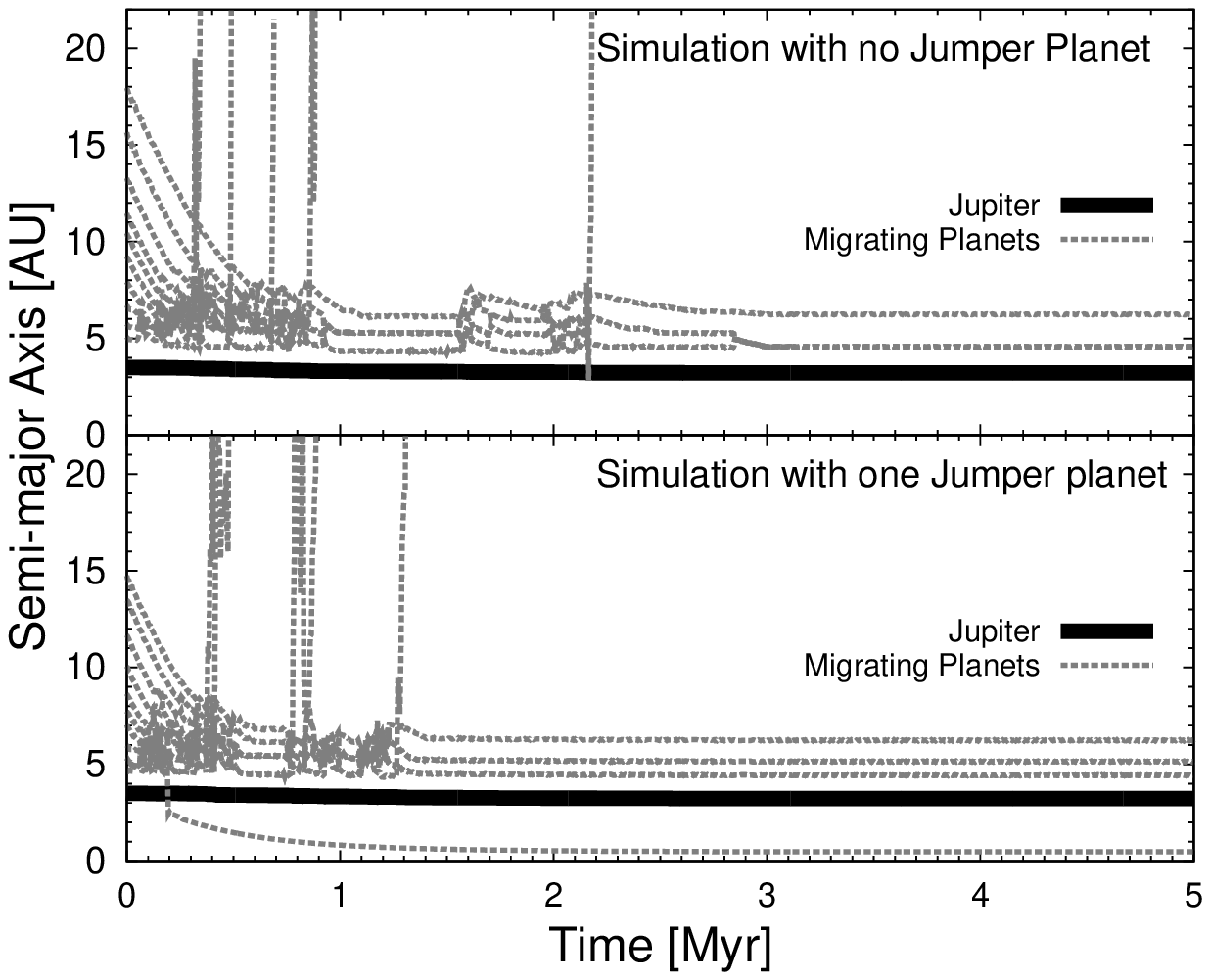}
\vspace{1cm}
\caption{Dynamical evolution of two simulations, each starting with 10 super-Earths (gray dashed curves) of $3 \mearth$ initially placed beyond the orbit of a Jupiter-mass planet at 3.5 AU (thick solid line). The simulations were in disk B, a standard (non-depleted) profile, which dissipated on a 3 Myr exponential timescale. The orbit of surviving bodies were followed for an additional 2 Myr. Vertical grey lines the scattering of a super-Earth (by the giant planet) onto a distant, often unbound orbit. In the upper simulation two super-Earths survived on orbits external to the giant planet. In the lower simulation three super-Earths survived exterior to the gas giant and one super-Earth "jumped" into the inner system. }
\end{figure}

\begin{figure}[!h]
\centering
\includegraphics[scale=1.3]{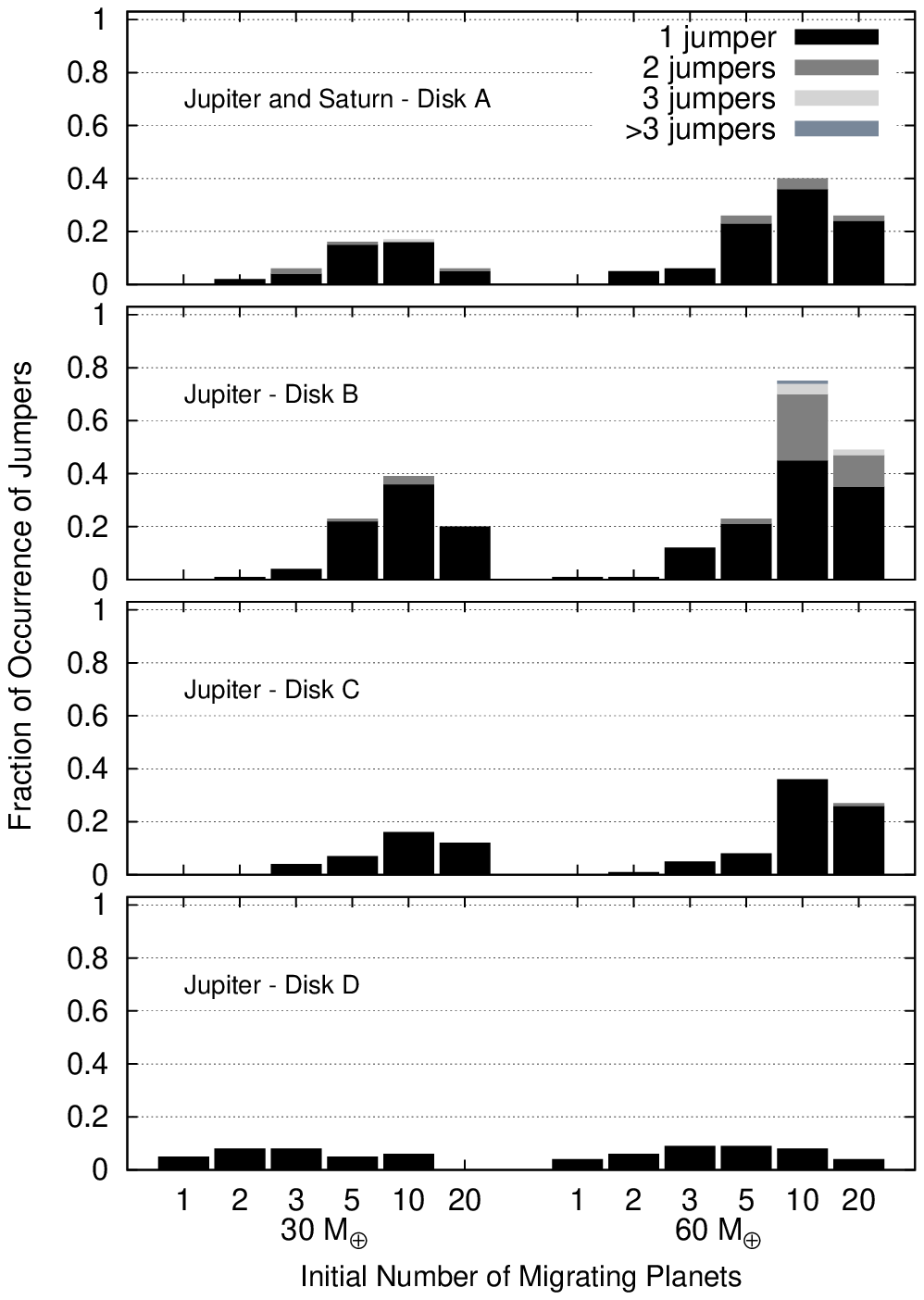}
\vspace{1cm}
\caption{The fraction of simulations that contain a jumper (i.e., the ``jump rate'') for the different gas-disk profiles.  Each histogram shows the jump rate as a function of the number of and total mass in super-Earths initially exterior to the giant planet(s).  Each jump rate is based on 100 different simulations with the same statistical setup.}
\end{figure}

\section{Discussion}

Our simulations tell a dynamical story about the origins of planetary systems.  The players in the story are a population of growing planetary cores at a few AU (e.g. beyond the snowline).  The cores are type I migrating inward.  If the cores grow slowly they migrate inward, pile up and the inner edge of the disk and form a compact system of many (perhaps 3-6 or even more) hot super-Earths (Cossou et al 2014), as observed (e.g., Lissauer et al 2011b; Fabrycky et al 2012).  

The story changes if we consider that some cores can grow sufficiently quickly to become gas giant planets.  If the innermost core grows into a gas giant then it blocks the inward migration of the other cores, which remain stranded in the outer disk and may potentially grow into ice- or gas-giants. Jumpers are more frequent in dense disks or in systems with many planetary cores and rarer in disks with few cores or inner cavities, such as disks undergoing photo-evaporation from the central star (Alexander \& Pascucci, 2012).

This story nicely places the Solar System in the context of extrasolar planetary systems.  If we assume that Jupiter was the innermost core and first to grow to a gas giant, it follows that Jupiter held back an invasion of inward-migrating bodies, notably Uranus, Neptune and perhaps Saturn's core. In the absence of Jupiter, the constituent cores of Saturn's core, Uranus and Neptune would have migrated into the inner solar system and become a system of multiple hot super-Earths. Migrating super-Earths would have perturbed or even prevented the formation of the Earth (Izidoro et al., 2014).  It therefore seems plausible that the presence of a Jupiter-like planet at few AU is an important factor for the formation of Earth-like planets.

What if a more distant planetary core becomes the first gas giant?  Any cores interior to the gas giant would simply be free to migrate inward. Embryos beyond the gas giant would be blocked in the outer system (except for the occasional jumper).  However, we expect the innermost core to generally grow the fastest (e.g. Lambrechts \& Johahsen, 2014) mainly because of the very strong dependence of accretion timescales on orbital radius.  In some situations a more distant core may indeed become a gas giant; in such cases close-in super-Earths should form but their total mass would be less than in a system with the same disk mass but no gas giants.

This model makes a clear observational prediction: the populations of planetary systems with close-in super-Earths and those hosting gas giant planets on Jupiter-like orbits should be anti-correlated.  Systems with many hot super-Earths (especially low-density super-Earths such as in the Kepler-11 system; Lissauer et al., 2011a) should not host a distant gas giant because such a planet should have acted as a migration barrier.  The in-situ accretion model for the origin of hot super-Earths makes the opposite prediction, that gas giants should be very common in systems with hot super-Earths (Schlaufman 2014).  

Systems with a {\it single} close-in super-Earth (Batalha et al., 2013; Lissauer et al., 2014) are unexpected because both formation models (in-situ formation and inward migration) suggest that hot super-Earths form in rich systems with many planets. Of course, many such systems may be false positives  or harbor additional not-yet-detected super-Earths (Hansen, 2013; Lissauer et al., 2014).  Our models suggests that systems which truly host just one super-Earth should also harbor a more distant giant planet. In other words, single super-Earths should be jumpers. This implies that the observed occurrence rate of single planet systems must be considerably lower than the occurrence rate of gas giants that may act as dynamical barriers. Radial velocity surveys estimate the frequency of gas giants with orbital periods shorter than 1000 days to be $\sim 15\%$ (Mayor et al 2011). Microlensing observations derive a much higher occurrence rate ($\sim 50\%$; Gould et al 2010). The frequency of single super-Earth systems is hard to measure given observational biases and the possibility of astrophysical false positives masquerading as single-planet systems.  Fang \& Margot (2012) estimated that half of all systems detected by Kepler contain a single planet.  Assuming a 30-40\% super-Earth occurrence rate, this equates to 15-20\% of stars hosting single super-Earths.  In contrast, Mayor et al. (2011) found a much higher ($> 70\%$) rate of multiplicity among super-Earth systems.  Given the complex inherent biases, we consider the occurrence rate of single super-Earths an open question.  The ratio of observed single-to-double super-Earth systems is also too uncertain to constrain our model at present, in part because two-planet systems may or may not represent jumpers depending on the circumstances. Of course, systems with just 1-2 super-Earths are also produced if it is not the innermost but the second- or third-innermost embryo that becomes a gas giant.  This prediction remains valid unless gas giants systematically grow from very distant planetary cores. We do not expect this to be the case.  Rather, formation models support our assumption that the innermost core should grow the fastest (Lambrechts and Johansen (2014); Levison et al., 2014 in preparation).  If our observational predictions are confirmed it would provide support for these formation models.

Of course, in our solar system rocky planets formed interior to the orbit of Jupiter. Thus,
even if the innermost embryo grows into a giant planet and acts as a perfect migration barrier, some disks may contain enough material close to their star to form systems of close-in (volatile-depleted) Earth-sized planets.  
In our solar system, rocky protoplanetary embryos inside the orbit of Jupiter were probably too small ($\sim$Mars-mass) to migrate substantially during the gas lifetime ($\tau_{gas}\ll\tau_{mig}$). To clearly identify systems that were not shaped by migration may thus require statistics of planets down below the threshold in mass for type-I migration ($\sim$Mars-mass).

As discussed before, the current collection of extra-solar planetary systems does not yet statistically evaluate these predictions.  Several systems appear to confirm our predictions: systems with many close-in super Earths that likely formed by inward migration (e.g. Kepler-11 and Kepler-32; Bodenheimer \& Lissauer, 2014; Swift et al., 2013); systems with a single transiting hot super-Earth (e.g. Kepler-22 and Kepler 67); a system with a super-Earth exterior to the orbit of a giant planet (Kepler 87; Ofir et al., 2014); and a system with a single hot super-Earth detected inside the orbit of a giant planet (GJ 832; Wittenmyer et al., 2014). However, other cases do not follow our predictions, such as the Kepler-90 and Kepler 48 systems that contain multiple super-Earths inside the orbits of giant planets.  What is needed for a definitive judgement is a large de-biased sample of planetary system structures, ideally with a rough indication of planetary composition (Marcy et al., 2014). This will become possible in the near future, particularly with the upcoming transit mission TESS and PLATO as well as improved radial velocity surveys (i.e. ESPRESSO). The confirmation or disproval of our predictions will be a crucial test for the scenario of origin of close-in super-Earths by radial migration.

\acknowledgments
We thank the anonymous referee for the very helpful and constructive comments. A~I. thanks financial support from CAPES foundation via grant 18489-12-5.  S. N. R, A. M., F. H. and A. P. thank the Agence Nationale pour la Recherche for support via grant ANR-13-BS05-0003 (project MOJO). We are very grateful to the CRIMSON team for managing the mesocentre SIGAMM where these simulations were performed.

\end{document}